\documentclass[preprint,3p]{elsarticle}

\usepackage{amssymb}
\usepackage{amsmath}
\usepackage{lipsum}

\usepackage{breqn}
\usepackage{braket}
\usepackage{xcolor}
\usepackage{graphicx}
\usepackage{slashed}
\newcommand{\as}{a_s}

\allowdisplaybreaks

\journal{Physics Letters B}

\begin{document}

\begin{flushright}
MSUHEP-23-028, ZU-TH 63/23
\end{flushright}
\vskip 3.5cm

\begin{frontmatter}




\title{The $N_f \,C_F^3$ contribution to the non-singlet splitting function \\ at four-loop order}

\author[1]{Thomas Gehrmann}
\ead{thomas.gehrmann@uzh.ch}

\author[2,3]{Andreas von Manteuffel}
\ead{manteuffel@ur.de}

\author[1]{Vasily Sotnikov}
\ead{vasily.sotnikov@physik.uzh.ch}

\author[1,3]{Tong-Zhi Yang}
\ead{toyang@physik.uzh.ch}

\affiliation[1]{organization={Physik-Institut},
            addressline={Universität Zürich}, 
            city={Winterthurerstrasse 190},
            postcode={8057 Zürich}, 
            country={Switzerland}}          
\affiliation[2]{organization={Institut für Theoretische Physik},
            addressline={Universität Regensburg}, 
            postcode={93040 Regensburg}, 
            country={Germany}}

\affiliation[3]{organization={Department of Physics and Astronomy},
            addressline={ Michigan State University},
            city={East Lansing},
            postcode={ MI 48824}, 
            country={USA}}

\begin{abstract}
We report a new result for the $N_f \,C_F^3$ contribution to the four-loop anomalous dimensions of non-singlet, twist-two operators in  Quantum Chromodynamics. This result is obtained through computations of off-shell operator matrix elements. Employing integration-by-parts reductions and differential equations with respect to a tracing parameter allowed us to derive analytic results valid for arbitrary Mellin moment $n$. 
\end{abstract}


\end{frontmatter}




\section{Introduction}
The theory predictions of high-energy hadronic collider observables rely on the factorization theorem in Quantum Chromodynamics (QCD), which states that the hadronic cross section is factorized into universal parton distribution functions (PDFs) and partonic cross sections. The partonic cross sections are perturbatively calculable. The PDFs are non-perturbative quantities, but their scale evolution (the well-known DGLAP evolution~\cite{Altarelli:1977zs,Gribov:1972ri,Dokshitzer:1977sg}) is governed by splitting functions, which can be evaluated perturbatively in QCD. 

Several benchmark partonic cross sections in QCD have been evaluated to next-to-next-to-next-to-leading order (N$^3$LO), see for example~\cite{Anastasiou:2015vya,Mistlberger:2018etf,Duhr:2020seh,Heinrich:2020ybq}. To achieve the same accuracy for hadronic cross sections, it is necessary to know the N$^3$LO PDFs, which require the knowledge of four-loop splitting functions. The splitting functions at three-loop accuracy in QCD were computed almost 20 years ago~\cite{Moch:2004pa,Vogt:2004mw}, and allowed the complete determination of NNLO PDFs. At four-loop order, results are available only for some specific color structures, including the leading-power $N_f$ contributions to all channels~\cite{Gracey:1994nn,Gracey:1996ad,Davies:2016jie}, the $N_f^2$ contribution to non-singlet splitting functions~\cite{Davies:2016jie}, leading color contribution to non-singlet splitting functions~\cite{Moch:2017uml}, and recently the $N_f^2$ contributions to pure-singlet~\cite{Gehrmann:2023cqm} and quark-to-gluon splitting functions~\cite{Falcioni:2023tzp}. Beyond these leading color, leading and sub-leading $N_f$ contributions, a finite number of Mellin moments were computed for all splitting functions  in~\cite{Moch:2017uml,Moch:2021qrk,Falcioni:2023luc,Falcioni:2023vqq,Moch:2023tdj}. Those results were already used to obtain approximate N$^3$LO PDFs~\cite{McGowan:2022nag,Hekhorn:2023gul}. 

The anomalous dimensions $\gamma(n)$ with Mellin moments $n$ are related to splitting functions $P(x)$ via the following Mellin transformation,
\begin{equation}
\label{eq:MellinT}
\gamma(n) =  - \int_0^1 d x x^{n-1} P(x)\,.
\end{equation} 
To go beyond the currently available all-$n$ results, in this paper, we consider one of the simplest remaining contributions: the $N_f \,C_F^3$ contribution to the non-singlet splitting functions, that also appear in Quantum Electrodynamics (QED). Following closely references~\cite{Gehrmann:2023ksf,Gehrmann:2023cqm}, we performed our computations in the framework of the operator product expansion (OPE), and extracted the splitting functions from the single pole of off-shell operator matrix elements (OMEs). The off-shell OMEs are defined as the off-shell matrix elements with an operator insertion, for the case of two partons in the external states it is 
\begin{align}
\label{eq:OMEsDe}
A^{}_{ij} = \braket{j(p)|O_i|j(p)}^{} \text{ with } p^2<0\,,
\end{align}
where $O_i$ is a twist-two operator. In the current context, it is the following quark non-singlet operator, 
\begin{equation}
\label{eq:nonsingletOP}
O^{}_{\text{ns}}(n) = \frac{i^{n-1}}{2}  \bigg[ \bar{\psi}_{i_1} \Delta\cdot \gamma (\Delta \cdot D)_{i_1 i_2} (\Delta \cdot D)_{i_2 i_3} \cdots (\Delta \cdot D)_{i_{n-1} i_n} \,\frac{\lambda_k}{2} \psi_{i_n} \bigg],\, k = 3, \cdots N_f^2-1 \,. 
\end{equation}  
In the above equation, $\lambda_k/2$ denotes diagonal generators of the flavor group $\text{SU}(N_f)$, and $\Delta$ is a light-like reference vector with $\Delta^2 =0$. The symbol $\psi$ represents the quark field, and $D^\mu_{ij} = \partial_\mu \delta_{ij} -i g_s  T_{ij}^a  A^a_\mu $ is the covariant derivative in the fundamental representation of a general gauge group. 

Compared with the conceptually complicated renormalization~\cite{Gross:1974cs,Dixon:1974ss,Hamberg:1991qt,Falcioni:2022fdm,Gehrmann:2023ksf} in the singlet sector, the non-singlet sector is much easier and allows for a straight-forward multiplicative renormalization: 
\begin{equation}
\label{eq:nsRenormalization}
O^{\text{R}}_{\text{ns}}(\mu, n)  = Z_{\text{ns}}(\mu, n) O^{\text{B}}_{\text{ns}}(n) \,,
\end{equation}
where superscripts B and R are used to represent the bare and renormalized operators, respectively. The renormalized operator satisfies the following renormalization group equation, 
\begin{equation}
\label{eq:evolutionOns}
\frac{d O_{\text{ns}}^{\text{R}}(\mu,n) }{ d \ln \mu} = -2 \gamma_{\text{ns}}(\mu, n) \,O_{\text{ns}}^{\text{R}}(\mu,n)\,,
\end{equation}
which defines the anomalous dimension $\gamma_{\text{ns}}$ of the non-singlet, twist-two operator. From equation~\eqref{eq:evolutionOns} and the fact of the bare operator $O_{\text{ns}}^{\text{B}}$ does not depend on the scale $\mu$, it is easy to see that 
\begin{equation}
\label{eq:renormalizaitonEqZ}
 \frac{d Z_{\text{ns}}(\mu,n) }{d \ln \mu} = -2  \gamma_{ \text{ns}}(\mu,n) \, Z_{\text{ns}}(\mu,n)  \,.
\end{equation}
It is not difficult to solve the above equation order by order in $a_s = \alpha_s/(4 \pi)$ with the help of the $d$-dimensional QCD $\beta$ function
\begin{align}
\label{eq:dBetafunction}
 \beta(\as,\,\epsilon) = \frac{d \as }{ d \ln \mu } = -2 \epsilon \, \as - 2 \as \sum_{i=0}^{\infty} \as^{i+1} \beta_i \,,
\end{align}
where $\epsilon = (4-d)/2$.
To four-loop order, the explicit result is given by
\begin{align}
\label{eq:ZfactorIntermsofGammaNS}
Z_{\text{ns}} = &\sum_{l=0}^\infty a_s^l Z_{\text{ns}}^{(l)} \nonumber \\ 
= & 1 + a_s \frac{\gamma^{(0)}_{\text{ns}}}{\epsilon}+a_s^2 \Bigg( \frac{\gamma_{\text{ns}}^{(1)}}{2 \epsilon} + \frac{1}{2 \epsilon^2} \bigg[ -\beta_0 \gamma_{\text{ns}}^{(0)} +  \big( \gamma_{\text{ns}}^{(0)}\big)^2  \bigg] \Bigg) \nonumber 
\\
& + a_s^3  \Bigg( \frac{1}{3 \epsilon} \gamma_{\text{ns}}^{(2)} + \frac{1}{6 \epsilon^2} \bigg[ -2 \beta_1 \gamma_{\text{ns}}^{(0)} - 2 \beta_0 \gamma_{\text{ns}}^{(1)}+ 3  \gamma^{(0)}_{\text{ns}} \gamma^{(1)}_{\text{ns}}    \bigg] \nonumber 
\\
& \quad + \frac{1}{6\epsilon^3} \bigg[ 2 \beta_0^2 \gamma_{\text{ns}}^{(0)}  - 3 \beta_0  \big( \gamma_{\text{ns}}^{(0)} \big)^2  + \big( \gamma_{\text{ns}}^{(0)} \big)^3   \bigg] \Bigg) \nonumber 
\\
&+ \frac{a_s^4}{24} \Bigg( \frac{1}{\epsilon^4} \bigg[-6 \beta _0^3 \gamma_{\text{ns}}^{(0)}-6 \beta _0 (\gamma _{\text{ns}}^{(0)})^3+11 \beta _0^2 (\gamma _{\text{ns}}^{(0)})^2+(\gamma
   _{\text{ns}}^{(0)})^4 \bigg] \nonumber 
   \\
   & \quad + \frac{1}{\epsilon ^3} \bigg[  {6 \beta _0^2 \gamma_{\text{ns}}^{(1)}-14 \beta _0 \gamma_{\text{ns}}^{(0)} \gamma_{\text{ns}}^{(1)}+12 \beta _0 \beta _1 \gamma_{\text{ns}}^{(0)}+6 \gamma_{\text{ns}}^{(1)} (\gamma
   _{\text{ns}}^{(0)})^2-8 \beta _1 (\gamma _{\text{ns}}^{(0)})^2}{} \bigg] \nonumber 
   \\
   & \quad +\frac{1}{\epsilon ^2}\bigg[{-6 \beta _0 \gamma_{\text{ns}}^{(2)}-6 \beta _1 \gamma_{\text{ns}}^{(1)}-6 \beta _2 \gamma_{\text{ns}}^{(0)}+8
   \gamma_{\text{ns}}^{(0)} \gamma_{\text{ns}}^{(2)}+3 (\gamma _{\text{ns}}^{(1)})^2}{}\bigg]+\frac{6 \gamma_{\text{ns}}^{(3)}}{\epsilon } \Bigg)+ 
 \mathcal{O}(a_s^5)\,,
\end{align}
where $\gamma_{\text{ns}}^{(l)}$ is defined as
\begin{equation}
\gamma_{\text{ns}}  =  \sum_{l=0}^\infty a_s^{l+1} \gamma_{\text{ns}}^{(l)}\,.
\end{equation}
Therefore, the four-loop, non-singlet anomalous dimension $\gamma_{\text{ns}}^{(3)}$ can be determined from the single pole in $\epsilon$ of the renormalization constant $Z_{\text{ns}}$. By separating the even and odd moments, $\gamma_{\text{ns}}$ can be decomposed as
\begin{align}
\label{eq:nsDecomposition}
    \gamma_{\text{ns}} = \frac{1+(-1)^n}{2} \gamma_{\text{ns}}^+ + \frac{1-(-1)^n}{2} \left(  \gamma_{\text{ns}}^- + \gamma^{\text{s}}_{\text{ns}} \right), 
\end{align}
where $\gamma_{\text{ns}}^{\text{s}}$ represents the flavour singlet but valence non-singlet contribution.

To extract the $Z_{\text{ns}}$ order by order in $a_s$, we insert equation~\eqref{eq:nsRenormalization} between two off-shell external quark states,
\begin{align}
\label{eq:2qomeForOqnsRenor}
     \braket{q|O^{\text{R}}_{\text{ns}}|q} &= Z_{q} \bigg[ Z_{\text{ns}} \braket{q|O^{\text{B}}_{\text{ns}}|q}^{\text{B}}  \bigg]\bigg|_{a_s^{\text{B}} \to Z_{a_s} a_s,\, \xi^{\text{B}} \to Z_g \xi } \,.
 \end{align}
Here we also need to consider the renormalization of the wave function, the strong-coupling constant, and the gauge parameter, all for $\xi=1$ in Feynman gauge. In addition to the explicit expressions for various contributions to $Z_q$, $Z_g$ and $Z_{a_s}$ that were collected in the appendix of the reference~\cite{Gehrmann:2023cqm}, we need only one more contribution to these renormalization constants, the $N_f \,C_F^3$ part of $Z_{q}^{(4)}$, which we document in appendix~\ref{sec:AppendixA}. In the following, we compute the $N_f \,C_F^3$ contribution to the four-loop corrections to the off-shell OME $\braket{q|O^{\text{B}}_{\text{ns}}|q}^{\text{B}}$.

\section{Methods and computations}
\begin{figure}
    \centering
    \begin{minipage}{0.45\textwidth}
           \includegraphics[scale=1.8]{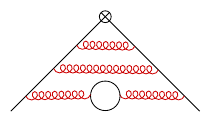} 
    \end{minipage} 
        \begin{minipage}{0.45\textwidth}
           \includegraphics[scale=1.8]{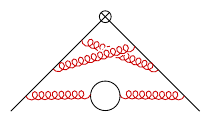} 
    \end{minipage} 
    \caption{Sample Feynman diagrams for the $N_f\,C_F^3$ contribution to the four-loop, non-singlet OME with two external quarks. The crossed circle represents the non-singlet operator $O_{\text{ns}}$.}
        \label{fig:diagrams}
\end{figure}

We generated the relevant Feynman diagrams with an insertion of the operator $O_{\text{ns}}$ by \texttt{QGRAF}~\cite{Nogueira:1991ex}; some sample diagrams can be found in Fig.~\ref{fig:diagrams}. The required Feynman rules for the operator $O_{\text{ns}}$ exhibit some peculiar patterns, i.e., terms like $(\Delta \cdot p)^{n-1}$ appear and thus prevent the application of standard integration-by-parts (IBP)~\cite{Chetyrkin:1981qh,Tkachov:1981wb,Laporta:2000dsw} algorithms in moment $n$-space. A method first proposed in~\cite{Ablinger:2012qm,Ablinger:2014nga} was used to overcome this difficulty, by summing these peculiar terms into linear propagators using a tracing parameter $t$. As an example, 
\begin{align}
\label{eq:sumXmethod}
( \Delta \cdot p)^{n-1} \to \sum_{n=1}^\infty t^n\, ( \Delta \cdot p)^{n-1} = \frac{t}{1- t \Delta \cdot p}\,.
\end{align}
After the desired manipulations have been performed, one can reexpand in $t$ to obtain the result for some moment $n$.
This method allows the applications of standard IBP algorithms and has been widely used to study the matching coefficient of heavy flavor quark contributions in deep-inelastic scattering~\cite{Ablinger:2014nga,Ablinger:2014vwa,Ablinger:2017tan,Behring:2019tus} and splitting function calculations~\cite{Blumlein:2021enk,Blumlein:2021ryt,Gehrmann:2023ksf} from off-shell OMEs.

We translated the Feynman rules in $n$-space to parameter $t$-space and then worked in $t$-space throughout. For this calculation, we used Mathematica to substitute the Feynman rules in $t$-space into the Feynman diagrams. \texttt{FORM}~\cite{Vermaseren:2000nd} was used to evaluate the traces of Dirac and color matrices. Regarding topology classification, we first used \texttt{Apart}~\cite{Feng:2012iq} and \texttt{MultivariateApart}~\cite{Heller:2021qkz} (see also \cite{Pak:2011xt,Bendle:2021ueg,Gerlach:2022qnc}) to eliminate the linear dependence among Feynman propagators. Then we classified all resulting topologies into integral families with 18 propagators by an in-house code. The code searches for the possible loop momentum transformations to see if one topology can be mapped into another one or not. To reduce the size of the unreduced amplitude, we first employed \texttt{Reduze2}~\cite{vonManteuffel:2012np} to eliminate integrals from zero sectors, and then applied shift relations to relate integrals between different sectors.

The IBP reductions for the amplitude were done by the private code \texttt{Finred}, which employs finite field sampling and rational reconstruction techniques~\cite{vonManteuffel:2014ixa,Peraro:2016wsq,Peraro:2019svx}. It is well-established that optimizing the input IBP system can significantly enhance the efficiency of the reduction process. In our study, we achieved this optimization by utilizing the linear algebra method~\cite{Agarwal:2020dye} in order to exert control over the generation of squared propagators. 

We used the differential equation (DE) method~\cite{Gehrmann:1999as} to determine the solutions of the master integrals. The derivation of the system of differential equations  for the master integrals follows the same approach as used for the amplitude reduction. In the first step, we select master integrals according to our generic integral ordering, and chose to ignore IBP relations generated from seeds in supersectors even if this leads to missing linear relations between the master integrals.
The rational functions in the differential equations are therefore somewhat complicated and involve rational numbers with many digits,
Consequently, for their reconstruction, we employed a significant number of samples for the variables and several finite fields, each with a cardinality of order $\mathcal{O}(2^{63})$. We chose to reveal these missing ``hidden'' relations by exploring the so-called scaling relations (see e.g.~\cite{Abreu:2022mfk}) of the master integrals. In our case, the scaling relations read
\begin{align}
\label{eq:rescaling}
    p^2 \frac{\partial  I_i }{\partial p^2} =  \frac{[I_i]}{2} \, I_i \,, 
\end{align}
where $I_i$ represents the $i$-th master integral and $[I_i]$ denotes mass dimension of $I_i$, and we set $[t]=0,\,[\Delta] =-1$ such that both $\Delta \cdot p$ and $t$ are dimensionless. In practice, through IBP reductions, the left-hand side of~\eqref{eq:rescaling} can not always be reduced to the simple form on the right-hand side. By enforcing the above scaling relations, we obtained several extra relations among the master integrals. Those additional relations greatly simplified the DE system. In the current case, the $N_f \,C_F^3$ contribution, we found 658 remaining master integrals, and the corresponding DE system with respect to $t$ can be cast into $\epsilon$-form~\cite{Henn:2013pwa} by applying the codes \texttt{CANONICA}~\cite{Meyer:2016slj,Meyer:2017joq} and \texttt{Libra}~\cite{Lee:2014ioa,Lee:2020zfb}. We obtained
\begin{equation}
    \mathrm{d}\vec{I}(t,\epsilon) = \epsilon \sum_i \mathrm{d}\ln(t-t_i) \mathbf{A}^{\!(i)} \vec{I}(t,\epsilon)\,,
\end{equation}
where we had set $p^2=-1$ and $\Delta \cdot p =1$. $\vec{I}$ is the vector of the new canonical master integrals, $\mathbf{A}^{\!(i)}$ are matrices involving rational numbers only, and $t_i=0,\pm 1, 2$. Interestingly, in addition to the letters appearing in harmonic polylogarithms (HPLs)~\cite{Remiddi:1999ew}, a new letter $t-2$ appears. This new letter enters the solutions of canonical master integrals starting from transcendental weight 7 only and does not contribute to the $N_f \,C_F^3$ contributions to the non-singlet splitting functions. It would be curious to see if the new letter contributes, e.g., to the finite part of corresponding off-shell OMEs or not. We leave it to future study. 

The boundary conditions can be conveniently chosen in the limit  $t \to 0$, where the linear propagators trivialize, and additional relations between the master integrals allow for their further reduction. The resulting master integrals are four-loop self-energy integrals~\cite{Baikov:2010hf,Lee:2011jt}, which in the present case were mapped to the master integrals for two-point functions in \cite{vonManteuffel:2019gpr,Lee:2023dtc}. By mapping to self-energy master integrals, we were able to determine the boundary conditions for all 658 master integrals. In practice, it is easier to first apply the regularity conditions: no branch cuts can be generated in the Taylor series of equation~\eqref{eq:sumXmethod}. We thus solved the canonical differential equations in the limit $t \to 0$ by keeping $\epsilon$ to all orders, and we set $c_a$ to zero for terms $c_a\,t^{\pm a \epsilon}$ ($a$ is a positive integer) in the resulting solutions.  In this way, we expanded the canonical solutions to transcendental weight 7 in terms of HPLs and generalized polylogarithms (GPLs) with the letter $t-2$.

The amplitude reduction was performed directly in terms of the canonical basis, where we can use the anticipated factorization of the $\epsilon$ and $t$ dependence in the denominators, and construct the denominators first~\cite{Abreu:2018zmy,Heller:2021qkz}. This helps to reduce the number of numerical samples required to reconstruct the functional dependence in $\epsilon$ and $t$. We subsequently inserted the solutions of the canonical basis into the amplitude and expanded the resulting amplitude order by order in $\epsilon$. We observed the emergence of Harmonic Polylogarithms (HPLs) with weights up to 6 in the single pole of the amplitude. For this $\epsilon$-expanded amplitude, we reconstructed also the rational numbers from their images in various finite fields. It is expected that the rational numbers appearing in the $\epsilon$-expanded amplitude are simpler, thus fewer finite fields are required for their reconstruction.

In this manner, we expressed the result for the $N_f \,C_F^3$ contribution to the bare four-loop OME, denoted as $\braket{q|O^{\text{B}}_{\text{ns}}|q}^{\text{B}}$, in terms of Harmonic Polylogarithms (HPLs) in parameter-$t$ space. Subsequently, we transformed this expression to $n$-space using harmonic sums~\cite{Vermaseren:1998uu,Blumlein:1998if}, aided by the \texttt{HarmonicSums} package~\cite{Ablinger:2009ovq,Ablinger:2012ufz,Ablinger:2014rba,Ablinger:2011te,Ablinger:2013cf,Ablinger:2014bra}. This transformation yielded the $n$-space representation for the bare OME.

\section{Results}
In the previous section, we obtained the $N_f \,C_F^3$ contribution to four-loop bare OME to the single pole in $\epsilon$ in $n$-space. The constant $Z_{\text{ns}}$ can be readily extracted from the renormalization equation~\eqref{eq:2qomeForOqnsRenor}. Notice that we also need the 3-loop corrections to the bare OME to order $\epsilon^0$, which was obtained previously for all color structures in~\cite{Gehrmann:2023ksf}. We checked explicitly that the obtained $Z_{\text{ns}}$ has the same form as in equation~\eqref{eq:ZfactorIntermsofGammaNS}, and thus allows the determination of the $N_f \,C_F^3$ contributions to $\gamma^{(3)}_{\text{ns}}$ from the single pole of $Z_{\text{ns}}$. The flavor singlet but valence non-singlet contribution $\gamma_{\text{ns}}^{\text{s}}$ in ~\eqref{eq:nsDecomposition} vanishes for the color structure $N_f \,C_F^3$, which allows us to write down our result in the following unified form,
\begin{equation}
     \gamma^{(3)\,, \pm}_{\text{ns}}(n)\big|_{\textcolor{blue}{N_f \,C_F^3}} =\frac{1\pm (-1)^n}{2} \gamma_{\text{ns}}^{(3)}(n)\big|_{\textcolor{blue}{N_f \,C_F^3}}\,,
\end{equation}
with
\begin{align}
\label{eq:anomalousN}
    \gamma^{(3)}_{\text{ns}}(n)&\big|_{\textcolor{blue}{N_f \,C_F^3}} =  (-1)^n \textcolor{magenta}{\bigg\{}  \left(\frac{448}{3 (n+1)}+\frac{448}{3 (n+1)^2}-\frac{448}{3 n}+\frac{448}{3 n^2}\right) S_{-4} \nonumber
    \\
    &+\left(-\frac{1184}{3 (n+1)}-\frac{320}{(n+1)^2}+\frac{640}{3
   (n+1)^3}+\frac{1184}{3 n}-\frac{448}{n^2}+\frac{128}{3 n^3}\right) S_{-3}\nonumber
    \\
    &+\left(\frac{480}{n+1}+\frac{544}{3 (n+1)^2}+\frac{128}{3 (n+1)^3}+\frac{64}{3
   (n+1)^4}-\frac{480}{n}+\frac{800}{3 n^2}-\frac{128}{n^3}+\frac{64}{3 n^4}\right) S_{-2}\nonumber
    \\
    &+\bigg(-\frac{1968}{n+1}-\frac{1376}{3 (n+1)^2}-\frac{464}{3
   (n+1)^3}+\frac{128}{(n+1)^4}-\frac{128}{(n+1)^5}+\frac{1968}{n}-\frac{4384}{3 n^2}+\frac{1136}{n^3}\nonumber
   \\
   &-\frac{1152}{n^4}+\frac{896}{3 n^5}\bigg) S_1+\bigg(-\frac{416}{3
   (n+1)}-\frac{224}{3 (n+1)^2}+\frac{64}{3 (n+1)^3}-\frac{320}{3 (n+1)^4}+\frac{416}{3 n}-\frac{224}{3 n^2}\nonumber
   \\
   &
   -\frac{64}{3 n^3}-\frac{320}{3 n^4}\bigg)S_2+\bigg(-\frac{32}{3 (n+1)}+\frac{64}{3 (n+1)^3}+\frac{32}{3 n}-\frac{64}{3 n^3}\bigg)
   S_3\nonumber
   \\
   &+\left(\frac{64}{n+1}+\frac{64}{(n+1)^2}-\frac{64}{n}+\frac{64}{n^2}\right) S_4+\bigg(-\frac{32}{n+1}-\frac{128}{(n+1)^2}+\frac{1088}{3
   (n+1)^3}+\frac{32}{n}-\frac{256}{n^2}\nonumber
   \\
   &-\frac{320}{3 n^3}\bigg) \zeta _3+\left(\frac{256}{n+1}+\frac{256}{(n+1)^2}-\frac{256}{n}+\frac{256}{n^2}\right) S_1 \zeta
   _3+\left(\frac{96}{n+1}+\frac{96}{(n+1)^2}-\frac{96}{n}+\frac{96}{n^2}\right) \zeta _4 \nonumber
   \\
   &+\left(\frac{128}{n+1}+\frac{128}{(n+1)^2}-\frac{128}{n}+\frac{128}{n^2}\right)
   S_{-2,-2}+\left(\frac{128}{3 n}-\frac{128}{3 (n+1)}\right) S_{-2,1}\nonumber
   \\
   &+\left(\frac{1984}{3 (n+1)}+\frac{1664}{3 (n+1)^2}-\frac{1280}{3 (n+1)^3}-\frac{1984}{3
   n}+\frac{2432}{3 n^2}-\frac{256}{3 n^3}\right) S_{1,-2}\nonumber
   \\
   &+\left(\frac{1280}{n+1}+\frac{1408}{3 (n+1)^2}+\frac{640}{3 (n+1)^3}-\frac{1280}{n}+\frac{2432}{3
   n^2}-\frac{1664}{3 n^3}+\frac{512}{n^4}\right) S_{1,1}\nonumber
   \\
   &+\left(-\frac{512}{3 (n+1)}-\frac{256}{3 (n+1)^2}-\frac{256}{3 (n+1)^3}+\frac{512}{3 n}-\frac{256}{3
   n^2}+\frac{256}{3 n^3}\right) S_{1,2}\nonumber
   \\
   &+\left(-\frac{128}{n+1}-\frac{128}{(n+1)^2}+\frac{128}{n}-\frac{128}{n^2}\right)
   S_{1,3}+\left(\frac{128}{n+1}+\frac{128}{(n+1)^2}-\frac{128}{n}+\frac{128}{n^2}\right) S_{2,-2}\nonumber
   \\
   &+\left(-\frac{512}{3 (n+1)}-\frac{256}{3 (n+1)^2}-\frac{256}{3
   (n+1)^3}+\frac{512}{3 n}-\frac{256}{3 n^2}+\frac{256}{3 n^3}\right) S_{2,1}\nonumber
   \\
   &+\left(-\frac{128}{n+1}-\frac{128}{(n+1)^2}+\frac{128}{n}-\frac{128}{n^2}\right)
   S_{3,1}+\left(-\frac{512}{n+1}-\frac{512}{(n+1)^2}+\frac{512}{n}-\frac{512}{n^2}\right)
   S_{1,1,-2}\nonumber
   \\
   &+\frac{1488}{n}-\frac{1488}{n+1}-\frac{1376}{n^2}-\frac{80}{(n+1)^2}+\frac{952}{n^3}+\frac{328}{(n+1)^3}-\frac{1120}{3 n^4}+\frac{656}{3
   (n+1)^4}+\frac{288}{n^5}\nonumber
   \\
   &+\frac{928}{3 (n+1)^5}-\frac{128}{n^6}-\frac{1024}{3 (n+1)^6} \textcolor{magenta}{\bigg\}}  
   + 512 S_{-6}+\left(-\frac{704}{n+1}-352+\frac{704}{n}\right) S_{-5}\nonumber
   \\
   &+\left(\frac{6016}{3 (n+1)}-\frac{64}{3 (n+1)^2}-272-\frac{6016}{3 n}+\frac{704}{n^2}\right)
   S_{-4}+\bigg(-\frac{1584}{n+1}-\frac{1696}{3 (n+1)^2}+\frac{1024}{3 (n+1)^3}\nonumber
   \\
   &-\frac{784}{3}+\frac{1584}{n}-\frac{5536}{3 n^2}+\frac{1280}{3 n^3}\bigg) S_{-3}\nonumber
   \\
   &+\left(\frac{656}{3
   (n+1)}-\frac{800}{3 (n+1)^2}-\frac{512}{(n+1)^3}+\frac{192}{(n+1)^4}-\frac{304}{3}-\frac{656}{3 n}+\frac{512}{3 n^2}-\frac{640}{3 n^3}+\frac{64}{n^4}\right) S_{-2}\nonumber
   \\
   &+\bigg(-\frac{656}{3
   (n+1)}-\frac{856}{3 (n+1)^2}-\frac{360}{(n+1)^3}-\frac{1760}{3 (n+1)^4}+\frac{2048}{3 (n+1)^5}+\frac{572}{9}+\frac{656}{3 n}-\frac{216}{n^2}\nonumber
   \\
   &+\frac{184}{3 n^3}-\frac{224}{3
   n^4}+\frac{256}{3 n^5}\bigg) S_1+\bigg(-\frac{88}{n+1}-\frac{72}{(n+1)^2}-\frac{96}{(n+1)^3}+\frac{544}{3 (n+1)^4}-\frac{638}{3}+\frac{88}{n}\nonumber
   \\
   &-\frac{568}{3 n^2}-\frac{224}{3
   n^3}+\frac{160}{3 n^4}\bigg) S_2+\left(\frac{1184}{3 (n+1)}+\frac{896}{3 (n+1)^2}+\frac{8}{3}-\frac{1184}{3 n}+\frac{448}{3 n^2}+\frac{128}{3 n^3}\right) S_3 \nonumber
   \\
   &+\left(-\frac{928}{3
   (n+1)}-\frac{32}{3 (n+1)^2}+136+\frac{928}{3 n}+\frac{160}{n^2}\right) S_4+\left(-\frac{64}{n+1}+\frac{928}{3}+\frac{64}{n}\right) S_5-\frac{256 S_6}{3} \nonumber
   \\
   &+\left(-\frac{920}{3
   (n+1)}-\frac{512}{3 (n+1)^2}-\frac{640}{3 (n+1)^3}-\frac{212}{3}+\frac{920}{3 n}-\frac{704}{3 n^2}+\frac{256}{3 n^3}\right) \zeta _3+256 S_{-3} \zeta
   _3 \nonumber
   \\
   &+\left(-\frac{256}{n+1}+640+\frac{256}{n}\right) S_{-2} \zeta _3+\left(-\frac{128}{(n+1)^2}+\frac{592}{3}+\frac{128}{n^2}\right) S_1 \zeta
   _3 \nonumber
   \\
   &+\left(-\frac{128}{n+1}+192+\frac{128}{n}\right) S_2 \zeta _3-\frac{128 S_3 \zeta _3}{3}-192 S_{-2} \zeta _4-120 \zeta _4\nonumber
   \\
   &+\left(-\frac{160}{n+1}+240+\frac{160}{n}\right) \zeta _5-320
   S_1 \zeta _5-\frac{1408 S_{-5,1}}{3}-\frac{256}{3} S_{-4,-2}\nonumber
   \\
   &+\left(\frac{512}{3 (n+1)}+1600-\frac{512}{3 n}\right) S_{-4,1}+256 S_{-4,2}-128 S_{-3,-3} \nonumber
   \\
   &+\left(\frac{128}{3
   (n+1)}+\frac{1024}{3}-\frac{128}{3 n}\right) S_{-3,-2}+\left(-\frac{7040}{3 (n+1)}+\frac{1024}{3 (n+1)^2}-160+\frac{7040}{3 n}-\frac{512}{3 n^2}\right) S_{-3,1}\nonumber
   \\
   &+\left(-\frac{1280}{3
   (n+1)}+\frac{640}{3}+\frac{1280}{3 n}\right) S_{-3,2}-\frac{896}{3} S_{-2,-4}+\left(-\frac{128}{n+1}+640+\frac{128}{n}\right) S_{-2,-3}\nonumber
   \\
   &+\left(\frac{896}{3 (n+1)}-\frac{896}{3
   (n+1)^2}-\frac{896}{3 n}-\frac{128}{n^2}\right) S_{-2,-2}+\bigg(\frac{6880}{3 (n+1)}+\frac{1088}{(n+1)^2}-\frac{1280}{3 (n+1)^3}+64\nonumber
   \\
   &-\frac{6880}{3 n}+\frac{6848}{3 n^2}-\frac{1280}{3
   n^3}\bigg) S_{-2,1}-512 \zeta _3 S_{-2,1}+\left(\frac{128}{3 (n+1)}+\frac{896}{3 (n+1)^2}-\frac{128}{3 n}+\frac{896}{3 n^2}\right) S_{-2,2}\nonumber
   \\
   &+\left(\frac{128}{n}-\frac{128}{n+1}\right)
   S_{-2,3}-128 S_{-2,4}-1408 S_{1,-5}+\left(\frac{2560}{3 (n+1)}+\frac{6656}{3}-\frac{2560}{3 n}\right) S_{1,-4}\nonumber
   \\
   &+\left(-\frac{8960}{3 (n+1)}+\frac{512}{(n+1)^2}+160+\frac{8960}{3
   n}-\frac{512}{n^2}\right) S_{1,-3}\nonumber
   \\
   &+\left(-\frac{1792}{3 (n+1)^2}+\frac{512}{3 (n+1)^3}+\frac{1760}{3}+\frac{1792}{3 n^2}-\frac{512}{3 n^3}\right) S_{1,-2}-512 \zeta _3
   S_{1,-2}\nonumber
   \\
   &+\left(\frac{320}{3 (n+1)}+\frac{112}{(n+1)^2}-\frac{192}{(n+1)^3}+\frac{640}{3 (n+1)^4}-\frac{320}{3 n}+\frac{112}{n^2}+\frac{192}{n^3}-\frac{128}{3 n^4}\right)
   S_{1,1}\nonumber
   \\
   &+\left(-\frac{896}{3 (n+1)^2}+\frac{128}{(n+1)^3}+\frac{880}{3}+\frac{896}{3 n^2}-\frac{128}{n^3}\right) S_{1,2}-256 \zeta _3 S_{1,2}\nonumber
   \\
   &+\left(\frac{512}{3 (n+1)^2}+320-\frac{512}{3
   n^2}\right) S_{1,3}+\left(\frac{1280}{3 (n+1)}-\frac{2624}{3}-\frac{1280}{3 n}\right) S_{1,4}-128 S_{1,5}\nonumber
   \\
   &-\frac{4736 S_{2,-4}}{3}+\left(\frac{512}{n+1}+\frac{7936}{3}-\frac{512}{n}\right)
   S_{2,-3}\nonumber
   \\
   &+\left(-\frac{1792}{3 (n+1)}+\frac{512}{3 (n+1)^2}+160+\frac{1792}{3 n}-\frac{512}{3 n^2}\right) S_{2,-2}\nonumber
   \\
   &+\left(-\frac{896}{3
   (n+1)^2}+\frac{128}{(n+1)^3}+\frac{880}{3}+\frac{896}{3 n^2}-\frac{128}{n^3}\right) S_{2,1}-256 \zeta _3 S_{2,1}\nonumber
   \\
   &+\left(-\frac{896}{3 (n+1)}+\frac{128}{(n+1)^2}+160+\frac{896}{3
   n}-\frac{128}{n^2}\right) S_{2,2}+\left(\frac{512}{3 (n+1)}-\frac{320}{3}-\frac{512}{3 n}\right) S_{2,3}\nonumber
   \\
   &-\frac{1792 S_{2,4}}{3}-\frac{3328
   S_{3,-3}}{3}+\left(\frac{256}{n+1}+512-\frac{256}{n}\right) S_{3,-2}\nonumber
   \\
   &+\left(-\frac{1792}{3 (n+1)}+\frac{640}{3 (n+1)^2}-64+\frac{1792}{3 n}-\frac{640}{3 n^2}\right)
   S_{3,1}+\left(\frac{512}{3 (n+1)}+\frac{704}{3}-\frac{512}{3 n}\right) S_{3,2}\nonumber
   \\
   &-384 S_{3,3}-\frac{1024 S_{4,-2}}{3}+\left(\frac{1024}{3 (n+1)}+\frac{448}{3}-\frac{1024}{3 n}\right)
   S_{4,1}-\frac{896 S_{4,2}}{3}-\frac{896 S_{5,1}}{3}\nonumber
   \\
   &-1024 S_{-4,1,1}+\frac{256}{3} S_{-3,1,-2}+\left(\frac{4096}{3 (n+1)}+256-\frac{4096}{3 n}\right) S_{-3,1,1}\nonumber
   \\
   &-\frac{512}{3}
   S_{-3,1,2}-\frac{512}{3} S_{-3,2,1}-256 S_{-2,-2,-2}+\left(-\frac{256}{3 (n+1)}-1280+\frac{256}{3 n}\right) S_{-2,1,-2}\nonumber
   \\
   &+\left(-\frac{1280}{n+1}-\frac{2560}{3
   (n+1)^2}+\frac{1280}{n}-\frac{3584}{3 n^2}\right) S_{-2,1,1}+\left(\frac{512}{3 (n+1)}-\frac{512}{3 n}\right) S_{-2,1,2}+256 S_{-2,1,3}\nonumber
   \\
   &-256 S_{-2,2,-2}+\left(\frac{512}{3
   (n+1)}-\frac{512}{3 n}\right) S_{-2,2,1}+256 S_{-2,3,1}+\frac{1024}{3} S_{1,-4,1}+\frac{256}{3} S_{1,-3,-2} \nonumber
   \\
   &+\left(\frac{1024}{3 (n+1)}-\frac{11008}{3}-\frac{1024}{3 n}\right)
   S_{1,-3,1}-\frac{2560}{3} S_{1,-3,2}-256 S_{1,-2,-3}\nonumber
   \\
   &+\left(-\frac{1024}{3 (n+1)}+\frac{1792}{3}+\frac{1024}{3 n}\right) S_{1,-2,-2}+\bigg(\frac{3584}{n+1}-\frac{1024}{3
   (n+1)^2}+320-\frac{3584}{n}\nonumber
   \\
   &+\frac{1024}{3 n^2}\bigg) S_{1,-2,1}+\left(\frac{2048}{3 (n+1)}-\frac{1280}{3}-\frac{2048}{3 n}\right) S_{1,-2,2}-256 S_{1,-2,3}+\frac{5120}{3}
   S_{1,1,-4}\nonumber
   \\
   &-\frac{12800}{3} S_{1,1,-3}+\frac{2560}{3} S_{1,1,4}+1024 S_{1,2,-3}-\frac{2560}{3} S_{1,2,-2}-\frac{1280}{3} S_{1,2,2}+\frac{1024}{3} S_{1,2,3}\nonumber
   \\
   &+512 S_{1,3,-2}-\frac{2560}{3}
   S_{1,3,1}+\frac{1024}{3} S_{1,3,2}+\frac{2048}{3} S_{1,4,1}+\frac{512}{3} S_{2,-3,1}+\frac{512}{3} S_{2,-2,-2}\nonumber
   \\
   &+\left(-\frac{1024}{3 (n+1)}-\frac{9728}{3}+\frac{1024}{3 n}\right)
   S_{2,-2,1}-\frac{2048}{3} S_{2,-2,2}+1024 S_{2,1,-3}-\frac{2560}{3} S_{2,1,-2}\nonumber
   \\
   &-\frac{1280}{3} S_{2,1,2}+\frac{1024}{3} S_{2,1,3}+\frac{1024}{3} S_{2,2,-2}-\frac{1280}{3} S_{2,2,1}+256
   S_{2,2,2}+\frac{1280}{3} S_{2,3,1}\nonumber
   \\
   &+\frac{2048}{3} S_{3,-2,1}+\frac{1024}{3} S_{3,1,-2}+\left(\frac{256}{3 (n+1)}-\frac{896}{3}-\frac{256}{3 n}\right) S_{3,1,1}+256 S_{3,1,2}+256
   S_{3,2,1}\nonumber
   \\
   &+\frac{512}{3} S_{4,1,1}+1024 S_{-2,1,1,-2}+\frac{8192}{3} S_{1,-3,1,1}-\frac{512}{3} S_{1,-2,1,-2}\nonumber
   \\
   &+\left(-\frac{2048}{n+1}-512+\frac{2048}{n}\right) S_{1,-2,1,1}+\frac{1024}{3}
   S_{1,-2,1,2}+\frac{1024}{3} S_{1,-2,2,1}+\frac{2048}{3} S_{1,1,-3,1}\nonumber
   \\
   &-\frac{2048}{3} S_{1,1,-2,-2}+5120 S_{1,1,-2,1}+\frac{4096}{3} S_{1,1,-2,2}-\frac{2048}{3} S_{1,2,-2,1}+\frac{512}{3}
   S_{1,3,1,1}\nonumber
   \\
   &+\frac{7168}{3} S_{2,-2,1,1}-\frac{2048}{3} S_{2,1,-2,1}-4096 S_{1,1,-2,1,1}-\frac{15802}{9 n}+\frac{15802}{9 (n+1)}+\frac{4628}{3 n^2}\nonumber
   \\
   &+\frac{86}{3 (n+1)^2}-\frac{2516}{3
   n^3}-\frac{340}{(n+1)^3}+\frac{1024}{3 n^4}-\frac{536}{3 (n+1)^4}-\frac{704}{3 n^5}\nonumber
   \\
   &-\frac{1280}{3 (n+1)^5}+\frac{224}{3 n^6}+\frac{1184}{3 (n+1)^6}-32 \,.
\end{align}
Here, we follow the notation used in~\cite{Gehrmann:2023cqm} and omit the argument $n$ of the harmonic sums defined by 
\begin{align}
\label{eq:HarmonicDefinition}
S_{\pm m_1, \,m_2,\,\cdots m_d}(n) &= \sum_{j=1}^{n} (\pm 1)^{j} j^{-m_1} S_{m_2,\,\cdots m_d}(j) \quad(m_i \in \mathbb{N}),\nonumber\\
S_\emptyset(n)&=1\,.
\end{align}
Our result contains harmonic sums up to weight 6. Unlike for the singlet anomalous dimensions we computed before, we notice that only two kinds of denominators $1/n$ and $1/(n+1)$ appear in the above equation. Furthermore, the coefficients of the first  power of $1/n$ and $1/(n+1)$ differ by a minus sign only, thus we can always write them as a single term, for example, 
\begin{align}
    \frac{256}{3(n+1)} - \frac{256}{3 n} = \frac{-256}{3n (n+1)} \,. 
\end{align}
 The above all-$n$ result in~\eqref{eq:anomalousN} is new. Evaluating the result for fixed $n$, we found full agreement with the fixed $n\leq 16$ results derived in~\cite{Moch:2017uml}. Moreover, the all-$n$ results for $\zeta_4$ and $\zeta_5$ terms have been derived in~\cite{Davies:2017hyl} and~\cite{Moch:2017uml}, respectively. For these results, we also found full agreement. 

Applying an inverse Mellin transformation to equation~\eqref{eq:MellinT}, we obtained the corresponding splitting functions in momentum fraction $x$-space. This is achieved with the help of the function {\it{InvMellin}} in the package \texttt{HarmonicSums}; alternatively, the method proposed in~\cite{Behring:2023rlq} could be used. The explicit expressions for the corresponding splitting functions are provided in the ancillary files.  

It is interesting to study the various limits of the splitting functions. In the limit $x \to 0$, the result is free of power divergences and reads
\begin{align}
  P^{(3)\,,+}_{\text{ns}}\big|_{N_f \,C_F^3}  =& -\frac{4}{9} \log(x)^5  -\frac{20}{9} \log(x)^4 + \left(64 \zeta _2-\frac{16}{3}\right) \log(x)^3+ \left( \frac{1840 }{3} \zeta _2+\frac{256 }{3} \zeta _3 -\frac{170}{3} \right) \log(x)^2 \nonumber \\
  & + \left( \frac{3592}{3}  \zeta _2 +\frac{2080 }{3} \zeta_3 +304 \zeta _4+\frac{500}{3} \right) \log(x) + \frac{256 }{3} \zeta_3 \zeta_2 +\frac{5032}{3}  \zeta _2+1120 \zeta _3\nonumber \\
  &-\frac{1928 }{3} \zeta_4 -\frac{304}{3}  \zeta_5 +\frac{2410}{9} + \mathcal{O}(x), 
  \\
  P^{(3)\,,-}_{\text{ns}}\big|_{N_f \,C_F^3}  =&  \frac{76}{45} \log(x)^5+ \frac{196}{9} \log(x)^4 + \left(\frac{1072}{9}-\frac{128 }{3} \zeta _2\right) \log(x)^3\nonumber
  \\
  &+ \left(-\frac{1744 }{3} \zeta _2 -\frac{320 }{3} \zeta _3 +\frac{2686}{3}\right) \log(x)^2 + \left(-\frac{2936 }{3} \zeta _2 -1248
   \zeta _3-416 \zeta _4+\frac{8756}{3}\right)\log(x) \nonumber
   \\
   &-\frac{832}{3} \zeta _3 \zeta _2-984 \zeta _2-1248 \zeta _3-\frac{776}{3}  \zeta _4-208 \zeta _5+\frac{29194}{9} + \mathcal{O}(x) \,.
\end{align}
For $P_{\text{ns}}^{(3),+}$, the double logarithmically enhanced terms proportional to $\log(x)^k$ with $k=5,4$ have been derived in reference~\cite{Davies:2022ofz}, and we find full agreement.

In the limit $x \to 1$, $P^{(3)\,,+}_{\text{ns}}\big|_{N_f \,C_F^3}$ and $P^{(3)\,,-}_{\text{ns}}\big|_{N_f \,C_F^3}$ are identical to next-to-leading power, and the result can be written in the form
\begin{align}
\label{eq:xto1limit}
    P^{(3)\,,+}_{\text{ns}}\big|_{N_f \,C_F^3} &\approx P^{(3)\,,-}_{\text{ns}}\big|_{N_f \,C_F^3}=A_4 \big|_{N_f \,C_F^3} \,\left[\frac{1}{1-x}\right]_+ +   B_4\big|_{N_f \,C_F^3} \, \delta(1-x)  \nonumber \\
    &+  C_{4}\big|_{N_f \,C_F^3} \log (1-x)+  D_{4}\big|_{N_f \,C_F^3} - A_4\big|_{N_f \,C_F^3}  + \mathcal{O}(1-x)\,,
\end{align}
where the plus distribution is defined as 
\begin{equation}
    \left[ \frac{1}{1-x} \right]_+ f(x) = \frac{1}{1-x} \Big( f(x)-f(1) \Big)
\end{equation}
for a continuous test function $f(x)$. Our explicit results for the coefficients in~\eqref{eq:xto1limit} read
\begin{align}
\label{eq:xto1Explicit}
    &A_4 \big|_{N_f \,C_F^3} = \frac{592 }{3} \zeta _3-320 \zeta _5+\frac{572}{9}\,, \nonumber \\
 &B_4\big|_{N_f \,C_F^3} = 224 \zeta _3^2-\frac{256 }{3}\zeta _2 \zeta _3 -308 \zeta _3+162 \zeta _2-204 \zeta _4+912 \zeta _5-\frac{6434 }{9} \zeta _6+32 \simeq 80.779482 \,,\nonumber
 \\
 & C_4\big|_{N_f \,C_F^3} = 256 \zeta _3-\frac{880}{3}\,, \nonumber
 \\
 & D_{4}\big|_{N_f \,C_F^3} = 80 \zeta _2-192 \zeta _3+\frac{464 \zeta _4}{3}-\frac{638}{3}\,.
\end{align}
We note that in this notation, the perturbative expansions of the coefficients are again defined in powers of $a_s$, that is,
\begin{align}
    A(a_s) = \sum_{l=1}^\infty a_s^l A_l\,,
\end{align}
and similarly for the other coefficients.
Interestingly, all results shown in equation~\eqref{eq:xto1Explicit} have been derived before, numerically for $B_4\big|_{N_f  C_F^3}$~\cite{Das:2019btv}, analytically for the others~\cite{Grozin:2018vdn,Henn:2019swt,vonManteuffel:2020vjv,Dokshitzer:2005bf,Basso:2006nk}. 
We find perfect agreement with the literature, thus providing another strong check of our all-$n$ result in~\eqref{eq:anomalousN}.

The coefficient of $\left[\frac{1}{1-x} \right]_+$ denoted by $A(a_s)$ determines the cusp anomalous dimension~\cite{Korchemsky:1987wg}, and we find that our result above agrees with the $N_f \,C_F^3$ contribution to four-loop cusp anomalous dimension in~\cite{Grozin:2018vdn,Henn:2019swt,vonManteuffel:2020vjv}. The coefficient of $\delta(1-x)$ denoted by $B(a_s)$ is called the virtual anomalous dimension. In~\cite{Das:2019btv}, the numerical result $B_{4}\big|_{N_f \,C_F^3} = 80.780\pm 0.005$ has been obtained, which agrees well with our analytic result shown above. The numeric result of~\cite{Das:2019btv}, the analytic four-loop collinear anomalous dimensions of~\cite{Agarwal:2021zft,vonManteuffel:2020vjv}, together with the soft-rapidity correspondence derived in~\cite{Li:2016ctv,Vladimirov:2016dll,Vladimirov:2017ksc}, allowed the numerical determination of the four-loop rapidity anomalous dimensions~\cite{Moult:2022xzt,Duhr:2022yyp}. With our new result for $B_{4}\big|_{N_f \, C_F^3}$ in~\eqref{eq:xto1Explicit} in hand, we obtained the $N_f \,C_F^3$ contribution to the four-loop rapidity anomalous dimension analytically. In the convention of~\cite{Moult:2022xzt}, it reads
\begin{align}
    \gamma^{\text{R}}_3\big|_{N_f \,C_F^3} =  40 \zeta _3^2-\frac{2212 \zeta _3}{9}-37 \zeta _4+\frac{800 \zeta _5}{3}+100 \zeta _6-\frac{21037}{216}\,.
\end{align}
Finally, it was conjectured in~\cite{Dokshitzer:2005bf,Basso:2006nk} 
that the all-order results of $C(a_s)$ and $D(a_s)$ can be written in terms of $A(a_s)$ and $B(a_s)$: 
\begin{align}
\label{eq:allorderCD}
    C(a_s) = \left[A(a_s)\right]^2 , \quad
    D(a_s) = A(a_s) \left[ B(a_s) + \frac{1}{2 a_s} \beta(a_s) \right]
\end{align}
where for $\beta(a_s)$ the limit $\epsilon \to 0$ of equation~\eqref{eq:dBetafunction} is implied.
Performing expansions for all-order results in the equation~\eqref{eq:allorderCD} to $a_s^4$, it reads (see also~\cite{Moch:2017uml})
\begin{align}
    C_4=  2 A_1 A_3 + A_2^2\,,\quad
    D_4 = \sum_{l=1}^3 A_l \left( B_{4-l} -  \beta_{3-l} \right)\,,
\end{align}
where the four-loop quantities $C_4$ and $D_4$ depend on $A$, $B$ and the $\beta$ function from lower-loop orders only. Our results verify the above conjecture for the color factor $N_f \,C_F^3$ explicitly.

\section{Conclusions}

We analytically computed the $N_f \,C_F^3$ contribution to the four-loop, non-singlet anomalous dimension for arbitrary Mellin moments $n$ for the first time. The method is based on the framework of the operator product expansion, through the computations of off-shell operator matrix elements. In contrast to the singlet case, the renormalization of the non-singlet contributions computed here is conceptually straight-forward.
We introduced a tracing parameter to replace symbolic exponents depending on $n$.
In this way, we were able to employ standard integration-by-parts reductions and the method of differential equations to perform the computations. The obtained result in Mellin space is quite simple and involves the denominators $1/n$ and $1/(n+1)$ only.
We successfully cross-checked our expressions with the expressions for fixed moments $n\leq 16$ in~\cite{Moch:2017uml}.
From the $n$-space result, the corresponding splitting function was obtained through an inverse Mellin transformation.
For the splitting functions, we discussed the limits $x \to 0$ and $x \to 1$.
The limit $x \to 1$ is particularly interesting and involves the cusp and virtual anomalous dimensions. The sub-leading power contributions have been known before, either numerically or analytically. Our result for the virtual anomalous dimension allowed us to derive the $N_f\,C_F^3$ contribution to the four-loop rapidity anomalous dimension analytically, which had been known only numerically before.

\section*{Acknowledgements}
We acknowledge the European Research Council (ERC) for funding this work under the European Union's Horizon 2020 research and innovation programme grant agreement 101019620 (ERC Advanced Grant TOPUP) and the National Science Foundation (NSF) for support under grant number 2013859. Note added: We thank Andreas Vogt for sharing a comparison of our results in the limit $x \to 1$  with his independent numerical determination of the $n_f C_F^3$ contributions to $B_4$, in which he found agreement to 7 significant digits.

\appendix

\section{The quark wave function renormalization constants}
\label{sec:AppendixA}
The quark wave function renormalization constant
\begin{align}
    Z_q = \sum_{l=0}^\infty Z_q^{(l)} a_s^l
\end{align}
is required to four-loop order~\cite{Chetyrkin:1999pq,Chetyrkin:2004mf}. The lower order results were collected in the appendix of reference~\cite{Gehrmann:2023cqm}, we do not repeat them here, and list only the additional contribution needed for this paper:
\begin{align}
Z_{q}^{(4)}\Big|_{\textcolor{blue}{N_f \,C_F^3}} = \frac{\frac{\xi ^2}{4}-\frac{\xi }{3}}{\epsilon ^3}+\frac{\frac{\xi }{2}+\frac{7}{8}}{\epsilon ^2} + \frac{8 \zeta _3-\frac{19}{6}}{\epsilon }\,.
\end{align}







\providecommand{\href}[2]{#2}\begingroup\raggedright\endgroup

\end{document}